\documentclass[%
	final,
	journal,
    comsoc,
	letterpaper,
	oneside,
	twocolumn,
	nofonttune,%
]{IEEEtran}%
\PassOptionsToPackage{usenames,dvipsnames,svgnames,x11names,table,prologue}{xcolor}%
\PassOptionsToPackage{hyphens}{url}%
\PassOptionsToPackage{nomessages}{fp}%
\ifCLASSINFOpdf
  \usepackage[pdftex]{graphicx}
\else
  \usepackage[dvips]{graphicx}
\fi
\ifCLASSOPTIONcompsoc
   \usepackage[caption=false,font=normalsize,labelfont=sf,textfont=sf]{subfig}
\else
   \usepackage[caption=false,font=footnotesize]{subfig}
\fi

\usepackage{stfloats}
\usepackage[english]{babel}%
\usepackage[utf8]{inputenc}%
\usepackage[babel,style=english]{csquotes}%
\usepackage{hyphsubst}%
\usepackage[%
	activate={true,%
	nocompatibility},%
	final,%
	tracking=true,%
	kerning=true,%
	spacing=true,%
	factor=1100,%
	stretch=10,%
	shrink=10%
]{microtype}%
\usepackage{setspace}%
\usepackage{xcolor}%
\definecolor{todonotecol}{RGB}{250,0,0}%
\usepackage{xparse}
\usepackage[%
	colorlinks=false,%
	urlcolor=black,%
	linkcolor=black,%
	citecolor=black,%
	filecolor=black,%
	breaklinks,%
	]{hyperref}%
\usepackage{url}%
\usepackage[%
	acronym,%
	nopostdot,%
	seeautonumberlist,%
	shortcuts,%
	section=chapter,%
	toc,%
]{glossaries}%
\glsaddkey*{url}
{}
{\glsentryurl}
{\Glsentryurl}
{\glsurl}
{\Glsurl}
{\GLSurl}
\glsaddkey*{longpltwo}
{}
{\glsentrylongpltwo}
{\Glsentrylongpltwo}
{\glslongpltwo}
{\Glslongpltwo}
{\GLSlongpltwo}
\newglossaryentry{}{%
	name={},%
	plural={},%
	description={},%
}%
\newacronym{dmz}{DMZ}{demilitarized zone}
\newacronym{it}{IT}{Information Technology}
\newacronym{ot}{OT}{operational technology}
\newacronym{opcua}{OPC UA}{Open Platform Communications Unified Architecture}
\newacronym{tsn}{TSN}{Time-Sensitive Networking}
\newacronym{scada}{SCADA}{supervisory control and data acquisition}
\newacronym{vpc}{VPC}{Virtual Process Controller}
\newacronym{plc}{PLC}{programmable logic controller}
\newacronym{udp}{UDP}{User Datagram Protocol}
\newacronym{pubsub}{Pub/Sub}{Publish/Subscribe}
\newacronym{vm}{VM}{Virtual Machine}
\newacronym{pc}{PC}{personal computer}
\newacronym{ptp}{PTP}{Precision Time Protocol}
\newacronym{ntp}{NTP}{Network Time Protocol}
\newacronym{splc}{Soft-PLC}{Software-PLC}
\newacronym{rte}{RTE}{Real-Time Ethernet}
\newacronym{osi}{OSI}{Open Systems Interconnection}
\newacronym{mac}{MAC}{Media Access Control}
\newacronym{e2e}{E2E}{end-to-end}
\newacronym{tcp}{TCP}{Transmission Control Protocol}
\newacronym{hmi}{HMI}{Human-Machine Interface}
\newacronym{kpi}{KPI}{Key Performance Indicator}
\newacronym{mes}{MES}{Manufacturing Execution System}
\newacronym{rami4.0}{RAMI 4.0}{Reference Architectural Model Industrie 4.0}
\newacronym{tacnet4.0}{TACNET 4.0}{Tactile Internet 4.0}
\newacronym{3gpp}{3GPP}{3rd Generation Partnership Project}
\newacronym{5gppp}{5GPPP}{5G Infrastructure Public Private Partnership}
\newacronym{itu}{ITU}{International Telecommunication Union}
\newacronym{ngnm}{NGNM}{Next Generation Mobile Networks}
\newacronym{agv}{AGV}{Automated Guided Vehicle}
\newacronym{v2v}{V2V}{vehicle-to-vehicle}
\newacronym{v2x}{V2X}{vehicle-to-infrastructure}
\newacronym{d2x}{D2X}{device-to-x}
\newacronym{qos}{QoS}{Quality of Service}
\newacronym{noa}{NOA}{NAMUR Open Architecture}
\newacronym{dcs}{DCS}{distributed control system}
\newacronym{m2m}{M2M}{machine-to-machine}
\newacronym{sdn}{SDN}{software-defined networking}
\newacronym{erp}{ERP}{enterprise resource planning}
\newacronym{ip}{IP}{Internet Protocol}
\newacronym{lte}{LTE}{Long Term Evolution}
\newacronym{5g}{5G}{5th Generation}
\newacronym{6g}{6G}{6th Generation Wireless Communication Systems}
\newacronym{harq}{HARQ}{Hybrid Automatic Repeat Request}
\newacronym{rat}{RAT}{radio access technology}
\newacronym{mc}{MC}{Multi-Connectivity}
\newacronym{fbmc}{FBMC}{Filter Bank Multicarrier}
\newacronym{gfdm}{GFDM}{Generalized Frequency Division Multiplexing}
\newacronym{sla}{SLA}{service-level agreement}
\newacronym{5qi}{5QI}{5G QoS Indicator}
\newacronym{bmbf}{BMBF}{German Federal Ministry of Education and Research}
\newacronym{nrt}{NRT}{non real-time}
\newacronym{irt}{IRT}{isochronous real-time}
\newacronym{rt}{RT}{real-time}
\newacronym{ar}{AR}{augmented reality}
\newacronym{wlan}{WLAN}{wireless local area network}
\newacronym{cpu}{CPU}{central processing unit}
\newacronym{ram}{RAM}{random-access memory}
\newacronym{ue}{UE}{User Equipment}
\newacronym{ran}{RAN}{radio access network}
\newacronym{bs}{BS}{base station}
\newacronym{cn}{CN}{Core Network}
\newacronym{epc}{EPC}{evolved packet core}
\newacronym{mec}{MEC}{mobile edge cloud}
\newacronym{rtt}{RTT}{Round-Trip Time}
\newacronym{vpn}{VPN}{virtual private network}
\newacronym{vpf}{VPF}{Virtual Process Control Function}
\newacronym{vdi}{VDI}{Association of German Engineers}
\newacronym{ipc}{IPC}{industrial pc}
\newacronym{rtos}{RTOS}{real-time operating system}
\newacronym{vpcmo}{VPC M\&O}{VPC Management \& Orchestration}
\newacronym{i/o}{I/O}{input /output}
\newacronym{fbd}{FBD}{Function Block Diagram}
\newacronym{ldd}{LDD}{Ladder Logic Diagram}
\newacronym{sfc}{SFC}{Sequential Function Chart}
\newacronym{il}{IL}{Instruction List}
\newacronym{st}{ST}{Structured Text}
\newacronym{os}{OS}{Operating System}
\newacronym{ir}{IR}{Inactive Resource}
\newacronym{msc}{MSC}{message sequence chart}
\newacronym{icps}{ICPS}{industrial cyber-physical system}
\newacronym{cps}{CPS}{Cyber-Physical System}
\newacronym{mtd}{MTD}{Moving Target Defense}
\newacronym{iot}{IoT}{Internet of Things}
\newacronym{iiot}{IIoT}{industrial Internet of Things}
\newacronym{iiot2}{IIoT}{Industrial IoT}
\newacronym{k8s}{K8s}{Kubernetes}
\newacronym{rkt}{rkt}{CoreOS Rocket}
\newacronym{dcos}{DC/OS}{Distributed Cloud Operating System}
\newacronym{ml}{ML}{Machine Learning}
\newacronym{ai}{AI}{Artificial Intelligence}
\newacronym{ict}{ICT}{industrial communication technology}
\newacronym{kvm}{KVM}{kernel-based virtual machine}
\newacronym{ie}{IE}{Industrial Ethernet}
\newacronym{ias}{IAS}{industrial automation system}
\newacronym{cots}{COTS}{commercial off-the-shelf}
\newacronym{slat}{SLAT}{Second Level Address Translation}
\newacronym{gui}{GUI}{graphical user interface}
\newacronym{l2}{L2}{Layer 2}
\newacronym{l3}{L3}{Layer 3}
\newacronym{sme}{SME}{small and medium-sized enterprise}
\newacronym{sba}{SBA}{Service Based Architecture}
\newacronym{xaas}{XaaS}{Everything-as-a-Service}
\newacronym{5gc}{5GC}{5G Core}
\newacronym{vnf}{VNF}{Virtual Network Function}
\newacronym{nfv}{NFV}{Network Functions Virtualization}
\newacronym{amf}{AMF}{Access and Mobility Management Function}
\newacronym{ng-ran}{NG-RAN}{Next Generation Radio Access Network}
\newacronym{lmf}{LMF}{Location management function}
\newacronym{pcf}{PCF}{Policy Control Function}
\newacronym{ausf}{AUSF}{Authentication Server Function}
\newacronym{udm}{UDM}{Unified Data Management}
\newacronym{upf}{UPF}{User Plane Function}
\newacronym{smf}{SMF}{Session Management Function}
\newacronym{udr}{UDR}{Unified Data Repository}
\newacronym{nef}{NEF}{Network Exposure Function}
\newacronym{nf}{NF}{Network Function}
\newacronym{nrf}{NRF}{Network Repository Function}
\newacronym{mano}{MANO}{Management and Orchestration}
\newacronym{c/r}{C/R}{Checkpoint/Restore}
\newacronym{criu}{CRIU}{Checkpoint/Restore In Userspace}
\newacronym{ppm}{PPM}{Parallel Process Migration}
\newacronym{pdu}{PDU}{Protocol Data Unit}
\newacronym{nic}{NIC}{Network Interface Card}
\newacronym{ftp}{FTP}{File Transfer Protocol}
\newacronym{scp}{SCP}{Secure Copy Protocol}
\newacronym{rcp}{RCP}{Remote Copy Protocol}

\glsdisablehyper
\usepackage[%
	backend=biber,%
	style=ieee,%
	isbn=false,%
	hyperref=true,%
	maxbibnames=99,%
	sorting=none,%
	natbib=true,%
	language=english,%
	defernumbers=true,%
	]{biblatex}%
\DeclareFieldFormat{sentencecase}{\csname bbx@colon@search\endcsname#1}

\usepackage{balance}
\usepackage{nameref}
\usepackage{soul}
\usepackage{flushend}
\usepackage{textcomp}
\usepackage{calc}
\usepackage{xkeyval}
\usepackage{multirow}
\usepackage{tabulary}
\usepackage{makecell}
\usepackage{pgfplots}
\usepackage{tikz}
\usepackage{textcomp} 
\usepackage{verbatim}

\pgfplotsset{ grid style={
   line width = 0.1pt
  }, compat=1.16}

\newcommand{\mytilde}{{\raise.17ex\hbox{$\scriptstyle\mathtt{\sim}$}}}

\newlength\textheighttemp%
\newlength\textwidthtemp%
\newlength\textheightstd%
\newlength\textwidthstd%
\newlength\textheightold%
\newlength\textwidthold%
\newlength\tempheight%
\newlength\tempwidth%
\SetProtrusion{encoding={*},family={bch},series={*},size={6,7}}
              {1={ ,750},2={ ,500},3={ ,500},4={ ,500},5={ ,500},
               6={ ,500},7={ ,600},8={ ,500},9={ ,500},0={ ,500}}
\SetExtraKerning[unit=space]
    {encoding={*}, family={bch}, series={*}, size={footnotesize,small,normalsize}}
    {\textendash={400,400}, 
     "28={ ,150}, 
     "29={150, }, 
     \textquotedblleft={ ,150}, 
     \textquotedblright={150, }} 
\SetTracking{encoding={*}, shape=sc}{40}
\makeatletter
\let\blx@rerun@biber\relax
\makeatother
\usepackage{algorithm}
\usepackage{algpseudocode}
\usepackage{newtxmath}
\usepackage{amsmath}
				\newcommand{\disablewr}[1]{#1}%
				\newcommand{\newcommanddisw}[3]{\newcommand{#1}[1]{\disablewr{\textcolor{#2}{#3}}}}%
\renewcommand{\disablewr}[1]{}%
\definecolor{todocol}{named}{red}
\newcommanddisw{\todo}{todocol}{ToDo: #1}%
\definecolor{migucol}{named}{purple}%
\newcommanddisw{\migucom}{migucol}{{@}comment: #1}%
\newcommanddisw{\miguhigh}{migucol}{#1}%

\definecolor{darecol}{named}{blue}%
\newcommanddisw{\darecom}{darecol}{{@}comment: #1}%
\newcommanddisw{\darehigh}{darecol}{#1}%

\begin{document}%
%
\title{%
Downtime Optimized Live Migration of Industrial Real-Time Control Services
\thanks{This research was supported by the German Federal Ministry for Economic Affairs and Climate Action (BMWK) within the project FabOS under grant number 01MK20010C. The responsibility for this publication lies with the authors. This is a preprint of a work accepted but not yet published at the 31st IEEE International Symposium on Industrial Electronics (ISIE). Please cite as: M. Gundall, J. Stegmann, M. Reichardt, and H.D. Schotten: “Downtime Optimized Live Migration of Industrial Real-Time Control Services”. In: 2022 31st IEEE International Symposium on Industrial Electronics (ISIE), IEEE, 2022.}
}%
%
%
\author{%
\IEEEauthorblockN{%
    Michael Gundall\IEEEauthorrefmark{1}, %
    Julius Stegmann\IEEEauthorrefmark{1}, %
    Mike Reichardt\IEEEauthorrefmark{1}, %
    and Hans D. Schotten\IEEEauthorrefmark{1}\IEEEauthorrefmark{2} %
    \\%
}%
\IEEEauthorblockA{%
    \IEEEauthorrefmark{1}German Research Center for Artificial Intelligence GmbH (DFKI), \\ Kaiserslautern, Germany \\%
    \IEEEauthorrefmark{2}Department of Electrical and Computer Engineering, Technische Universität Kaiserslautern, \\ Kaiserslautern, Germany %
	\\%
    Email: %
        \{michael.gundall, julius\_raphael.stegmann, mike.reichardt, 
        hans\_dieter.schotten%
        \}@dfki.de
        \\%
}%
}%


%

%
%
%
%
%
%
%
%
\maketitle
\renewcommand{\figurename}{Fig.}
\renewcommand{\tablename}{Tab.}
%
%
%
\begin{abstract}%
Live migration of services is a prerequisite for various use cases that must be fulfilled for the realization of Industry 4.0. In addition, many different types of services need to provide mobility and consequently need to be migrated live. These can be offloaded algorithms from mobile devices, such as unmanned vehicles or robots, security services, communication services or classic control tasks. In particular, the latter place very high demands on determinism and latency. Here, it is of utmost importance that the downtime of the service is as low as possible. Since existing live migration approaches try to optimize multiple metrics such as downtime, migration time as well as energy consumption, which are equally relevant in the IT domain, it is not possible to use any of these approaches without adoptions.

Therefore, a novel concept is proposed that builds on top of both existing migration approaches as well as virtualization technologies and aims primarily at minimizing service downtime. Furthermore, the concept is evaluated using a test environment. The results show that a sub-millisecond downtime can be achieved with the proposed concept. Moreover, the total migration time is in the range of several hundred milliseconds for the highest performance setting and two seconds for a non-invasive approach.

\end{abstract}%
%
%
%
%
%
\IEEEpeerreviewmaketitle
%
%
%
%
%
%
%
%

\section{Introduction}%
\label{sec:Introduction}
Digitalization enables opportunities and benefits for a broad range of applications and use cases \cite{hozdic2015smart,etfa2018}. Furthermore, in manufacturing and process industries the Industry 4.0 vision was introduced \cite{lasi2014industry}. It describes scenarios and goals for highly digitized future factories. These goals are a shortened "time to market", increased resource efficiency, and customization of products \cite{lasi2014industry}. Looking deeper into the individual goals, each of them requires a very high flexibility of the factory modules. In order to increase the flexibility new, predominantly wireless, communication standards such as \gls{5g} \cite{access2021} and \gls{6g} \cite{jiang2021road} as well as powerful data processing tools are investigated. 

Moreover, so-called \glspl{agv} will increasingly move through industrial plants. Here, the offloading of algorithms and the interaction between different \glspl{agv} or \gls{agv} and factory modules require an increased networking between these devices. Very often, also stringent demands on \gls{qos}, such as latency, are required. Thus, the offloaded data has to be processed at servers close to the devices, so-called edge servers, instead of transferring it through the whole \gls{it} backbone to central data servers. This results in more and more distributed and networked systems. 

Hence, distributed systems will play a decisive role in the future, where it can often make sense to migrate the associated processes from one to another processing node. If the productivity should not be reduced, this should be done "live" that means without interruption of the normal operation of the plant. In this way, highly flexible, resilient, and low-latency applications can be realized. How this can ideally work without affecting its underlying functionality is to be investigated, with a focus on industrial control processes.

Accordingly, the following contributions can be found in this paper:
\begin{itemize}
    \item Analysis and assessment of existing live migration approaches for application in industrial applications
    \item Introduction of a novel downtime optimized live migration concept especially dedicated to fulfill industrial requirements
\end{itemize}

Therefore, the paper is structured as follows: 
Sec. \ref{sec:Migration Goals} lists different situations where live migration is required, while Sec. \ref{sec:Metrics} introduces 
metrics that are related to live migration in industrial environments. Further, a short description of existing live migration concepts and related work on this topic can be found in Sec. \ref{sec:Live Migration Concepts} and Sec. \ref{sec:Related Work}, respectively. Then, our improved concept is proposed (Sec. \ref{sec:Live Migration Concepts}) and evaluated (Sec. \ref{sec:Testbed and Evaluation}). Finally, a conclusion is given (see Sec. \ref{sec:Conclusion}).

\section{Migration Goals}%
\label{sec:Migration Goals}

Here it can be useful to transfer a process to another compute node for various reasons. Therefore, Tab. \ref{tab:goals} lists the most important use cases and the corresponding goals that are relevant for industrial applications. 
 
\begin{table*}[tb]
\caption{Overview of different migrations use cases, corresponding migration goals and sources of related investigations}
\label{tab:goals}
\centering
\begin{tabular*}{.999\textwidth}{|l|p{0.45\textwidth}|l|}
\cline{1-3}
\bfseries Migration use cases & \bfseries Migration goals & \bfseries Sources \\
\cline{1-3}
Efficiency, energy consumption &  Reduced communication power, reduced computation power & \cite{strunk2012costs,govindaraj2018container}\\
Interruption-free hardware replacement & Change of hardware, planned maintenance cycles, hardware upgradeste & \cite{strunk2012costs,goldschmidt2018container}\\
Availability, resilience, self-healing & Reaction to decreased QoS, reaction to decreased security level & \cite{icit2021,10.1007/978-3-642-36883-7_19}\\
Mobility support & Mobility of moving unmanned vehicles, movement of persons with mobile devices & \cite{govindaraj2018container,icit2021, Mobilkom2021} \\
(Seamless) reconfiguration and program updates & Lot size one, increase of product quality, reaction to decreased security level & \cite{8502526,goldschmidt2018container, icit2021, 10.1007/978-3-642-36883-7_19}\\
\cline{1-3}
\end{tabular*}
\end{table*}

The first use case that is well-known in \gls{it} data centres, is the migration of a process for energy savings and cost efficiency. As the load on the different servers could change, it can be beneficial to orchestrate the processes in this way that, for example, one of the servers could be spared. Therefore, both energy and money could be saved.

Next, it is possible that hardware has to be changed or replaced. Typically, this is done during  a specific maintenance phase, however, replacement of hardware components could also be required during operational phase. This is especially true, if parts break that do not directly cause a failure of the process but are related to it. Then, the process should be migrated to a different hardware node and the hardware can be replaced.

Further, preventing a failure of the process is a reason why a migration should be performed.  If any error occurs, such as \gls{qos} degradation or as security measure for compromised systems, a reallocation of one or more processes can also be required.

The same counts for mobile devices, such as \glspl{agv}. If, for example, such modules move between different production halls with different IT infrastructures, it is necessary to migrate the associated processes to a different computation node. In this way, optimal \gls{qos}, such as minimal latency, can always be achieved.
Last, live migration enables not only the redeployment of processes but also reconfiguration and rolling updates. If massive flexible production should be realized, with lot size one as most extreme example, this use case is of very high interest. However, live migration concepts are only part of seamless program updates, since also convergence criteria have to be investigated in detail.

\section{Metrics}%
\label{sec:Metrics}
In this section, most important metrics are to be defined. Here, the focus lies on services that are suitable for industrial environments. Further, effects on the process or application itself as well as on the executing infrastructure must be taken into account. Since \cite{strunk2012costs} already describes most important metrics, they are also the basis for this work.

\subsection{Process Downtime} The process downtime $T_D$ is the most important property for the process itself, especially for \gls{rt} applications. Here, the sum of the inactive time intervals of the process during the migration must be considered, since even many short interruptions can have serious effects on the \gls{qos}. Even more important, is the longest continuous interval of inactivity. This determines whether, for example, a real-time criterion can be met. For scheduled processes that are required for isochronous control tasks having a fixed sampling time, the triggering of the migration immediately after a CPU processing phase must also be taken into account. 

\subsection{Migration time} 

The migration time $T_M$ also plays an essential role. The longer the migration time, the more computation and communication resources are used and blocked for other processes. In addition, this metric is the main indicator for the maximum flexibility that could be reached. A process could not migrated more often than this value. This is important, if a mobile device, such as a drone, is changing factory halls in very short intervals.

\subsection{Transmitted data volume} 

The transmitted data volume $D$ is specified as the amount of data that has to be transferred during the whole migration phase. This value might be less important for one small data set and empty networks, however, network capacity in industrial environments is limited. In most of the cases, a real industrial network is heavily loaded. The more processes have to be migrated in parallel, the more important this parameter becomes.

\subsection{Resource load} 

In addition to communication system, the resource load $L$ of the host systems must also be considered. Furthermore, an increase of this value also increases the overall energy consumption. Especially in the area of edge and fog computing, where the computational resources are higher compared to field level, but still limited, the relationship between costs and benefits of virtualisation and live migration of processes plays a decisive role. 

\section{Existing Live Migration Approaches}%
\begin{figure}[tb]
\centering
 \subfloat[\gls{c/r} migration with inter-copy memory transfer.]{\resizebox{\columnwidth}{!}{%
\includegraphics{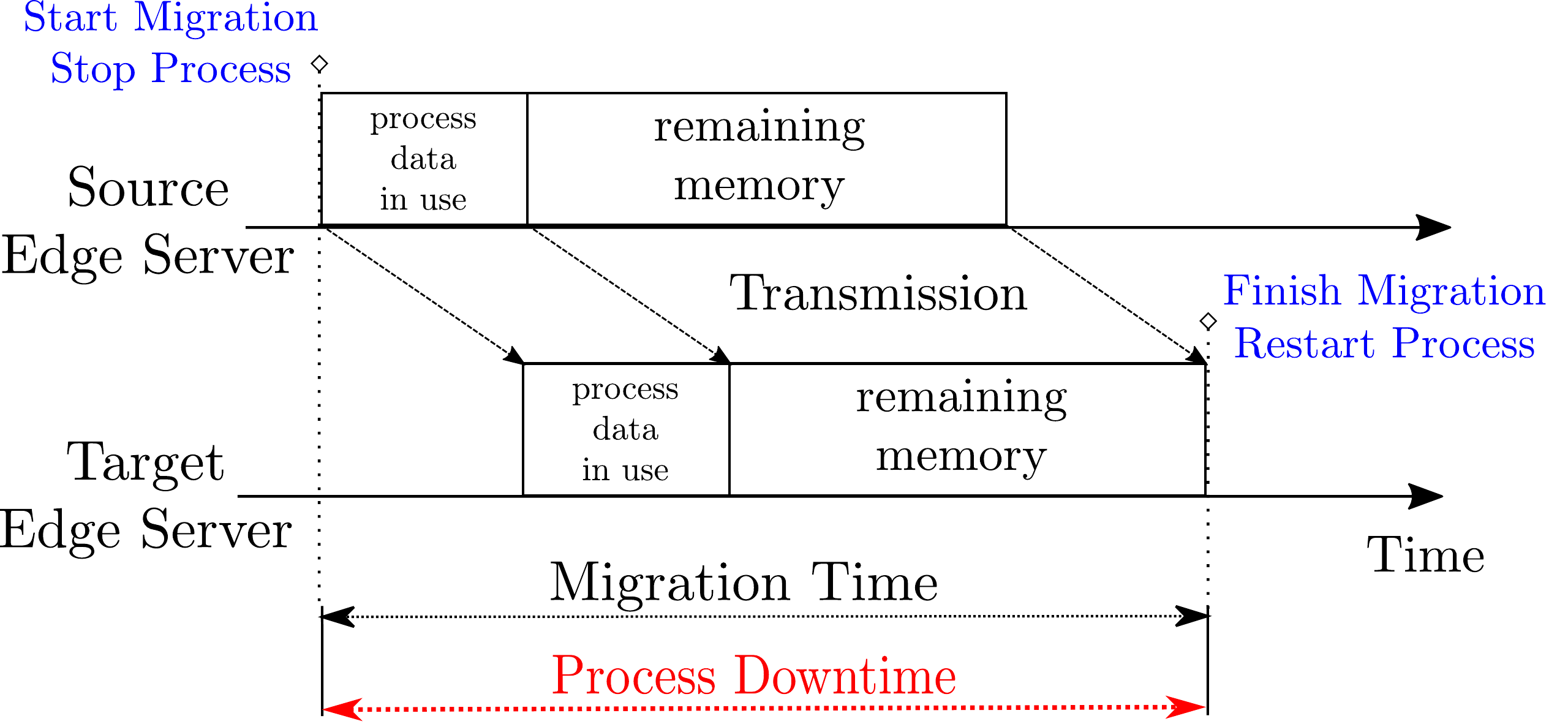}}\label{fig: inter copy}} 

\subfloat[\gls{c/r} migration with pre-copy memory transfer.]{\resizebox{\columnwidth}{!}{%
\includegraphics{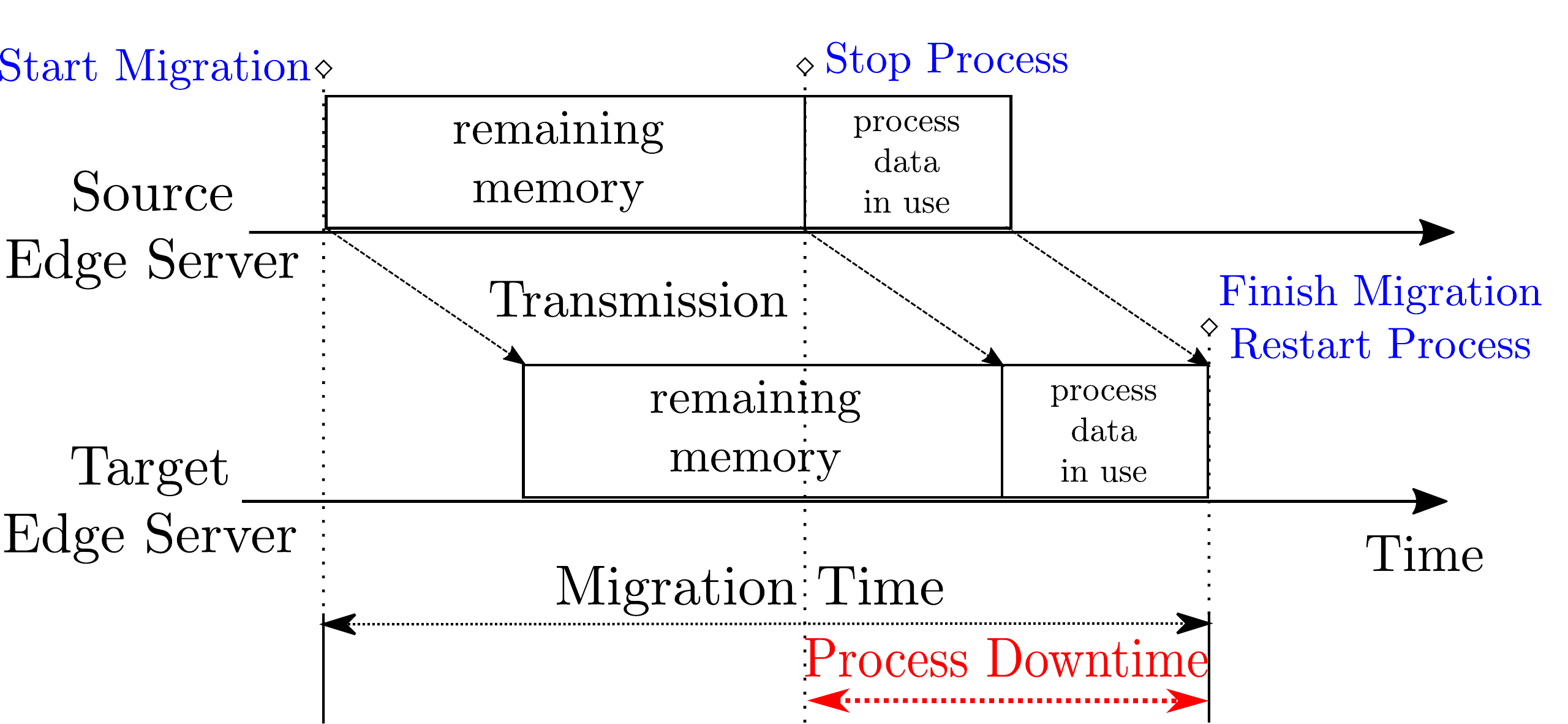}}\label{fig: pre copy}}

 \subfloat[\gls{c/r} migration with post-copy memory transfer.]{\resizebox{\columnwidth}{!}{%
\includegraphics{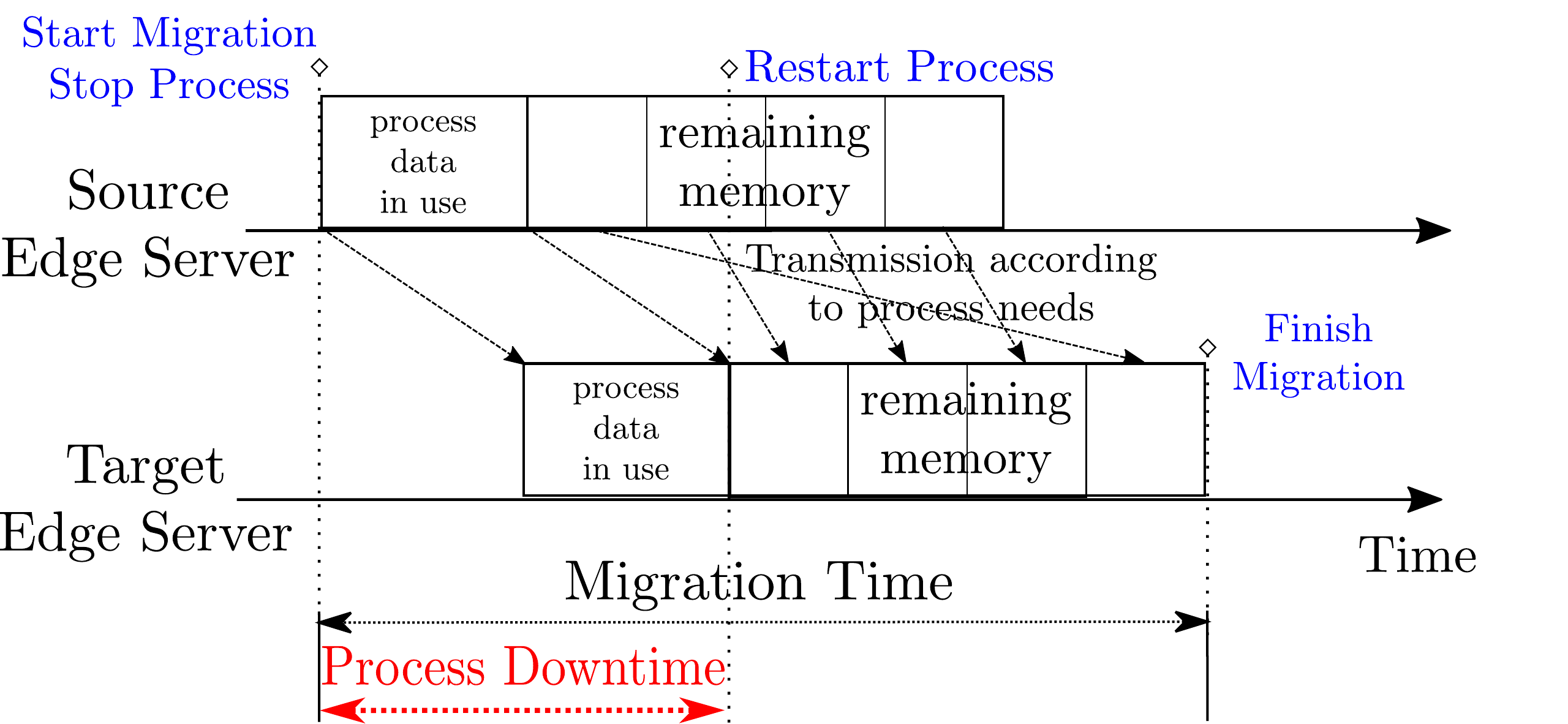}}\label{fig: poat copy}} 

\subfloat[PPM procedure including handover mechanism.]{\resizebox{\columnwidth}{!}{%
\includegraphics{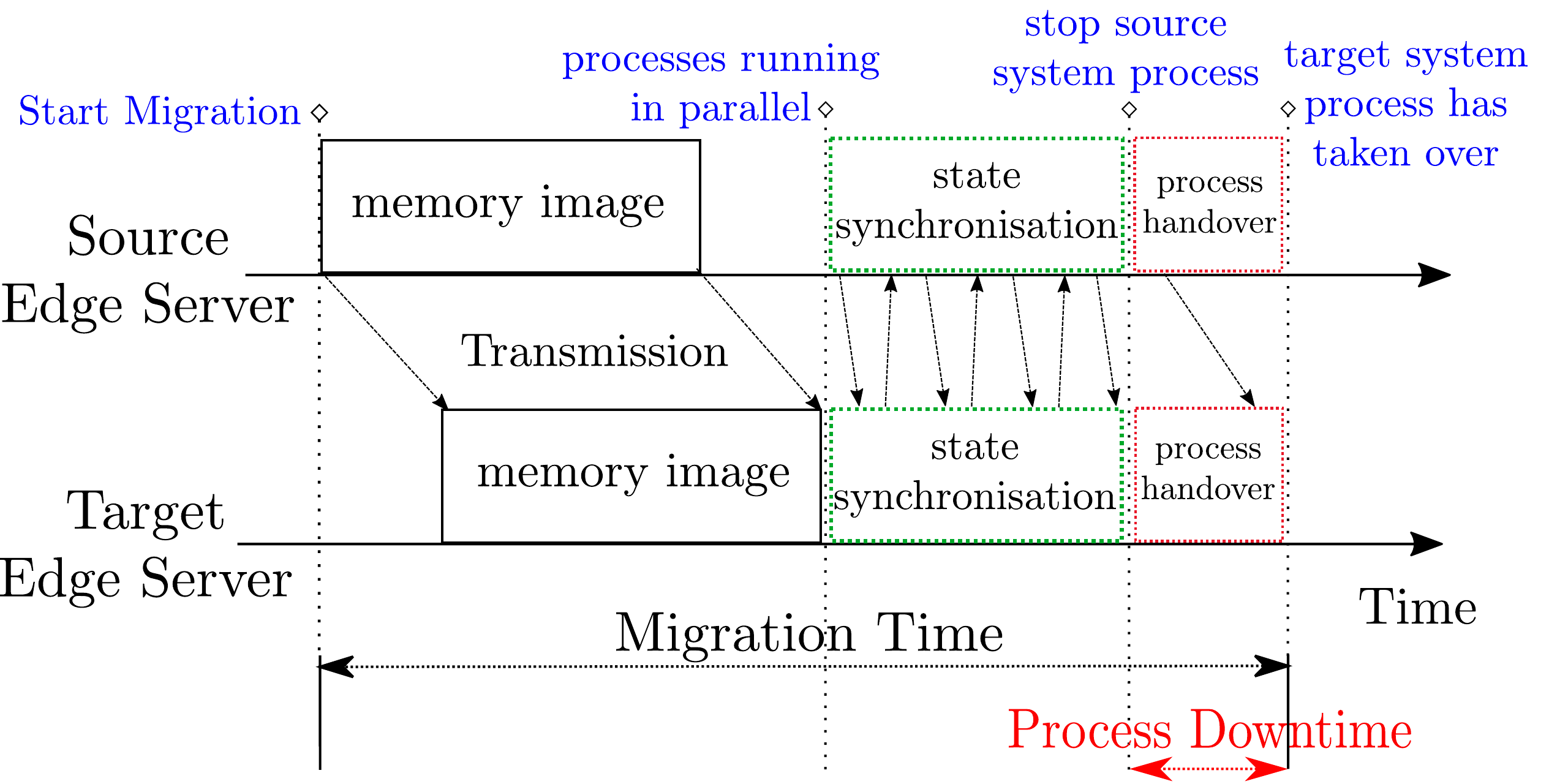}}\label{fig: parallel process}}
  \caption{Overview of the most common live migration approaches that are mostly used in \gls{it} environments until now \cite{Mobilkom2021}.}
  \label{fig:live migration approaches}
\end{figure}

\label{sec:Live Migration Concepts}
This section presents the current state of the art in terms of live migration, which is mostly applied in the \gls{it} environment nowadays. Here, the \gls{c/r} method is widely used. As the name indicates, the process is frozen and its status on the disk is checkpointed. Next, the data is transferred to a new target system and the process can be resumed. Recently, there is a development towards user-space-based approaches, since advantages of high transparency in combination with 
small invasive intervention in the central components of the operating system can be reached. In order to optimize specific metrics, several main \gls{c/r} approaches evolved. These are shown in Fig. \ref{fig:live migration approaches} and discussed in the following part.

\subsection{Inter-copy C/R}

The classical (inter-copy) \gls{c/r} procedure, which is shown in Fig. \ref{fig: inter copy}, consists of the following steps. Firstly, the process is stopped and all data in the memory is completely transferred from one system to another. Then, the process is resumed on the target system. Accordingly, for this procedure, it can be assumed that $T_M = T_D $.

\subsection{Pre-copy C/R}
Fig. \ref{fig: pre copy} shows the pre-copy tactic that aims at reduced process downtime. To achieve this goal, as much data as possible is transferred to the target system, before the process is frozen. Here, primarily data that is not expected to be needed in the coming process iterations is selected. Then the process is shut down on the source system, the remaining data is transferred, and the process is restarted on the target system.

\subsection{Post-copy C/R}
With the post-copy tactic on the other side, which is depicted in Fig. \ref{fig: poat copy}, the process is directly frozen at the beginning of the migration process, comparable to inter-copy \gls{c/r}. Afterwards, however, only the parts of the memory that are important for the next process iterations are transferred. The remaining parts of the memory are then transferred after the process has already restarted on the target system \cite{reber1}. This approach aims at a shorter migration time, but has a higher process downtime compared to pre-copy tactic. 
Both strategies are part of intensive research \cite{performance1,precopy1,hines2009post}. The post-copy strategies in particular increase the risk of a complete process failure if missing data cannot be transferred in time afterwards. The pre-copy strategy brings few advantages in terms of downtime if large parts of the data change in just a few process steps. Both methods require additional precise prediction of future steps.

\subsection{Hybrid C/R}
Hybrid \gls{c/r} is a combination of pre- and post-copy tactics. Here, most likely required memory pages are transferred to the target system before the process is stopped and transferred. Then the process is already resumed on the target system before the remaining memory pages are transferred. This approach is a good choice if both migration time and process downtime should be minimized, since the process downtime is only slightly higher compared to pre-copy and the migration time is only slightly higher compared to te post-copy tactic, and vice versa. with all the aforementioned approaches, however, the process downtime is still too high for industrial services, where mostly only outages of $<$1ms are acceptable.

\subsection{\gls{ppm}}

Therefore, latest approaches go one step further and use the parallel migration methodology \cite{parallel1,parallel2} that is also referred to as \gls{ppm} \cite{Mobilkom2021}. In previous approaches, only one instance of the process was active at a time. Thus, Fig. \ref{fig: parallel process}, 
describes the idea that the process is already running on the target system while both processes are supplied with the same data. When a migration is triggered, only a very small part of the memory has to be transferred to the target system. This results in significantly decreased process downtime  \cite{parallel1,parallel2}. Nevertheless, there are several key challenges to ensure a smooth handover, such as time and state synchronization, while also ensuring that all instances of the processes running in parallel are always supplied with the same data at the same time. 

\subsection{Analysis}
Each of the proposed live migration approaches has its benefits in one category and corresponding drawback in another. Thus, Tab. \ref{tab:comparison} maps the aforementioned metrics to each of the approaches. 
\begin{table}[tb]
\caption{Qualitative comparison between each of the proposed migration approaches and the corresponding metrics.}
\begin{center}
\begin{tabular}{|c|c|c|c|c|}
\cline{1-5}
\textbf{Migration} & \multicolumn{4}{|c|}{\textbf{Metrics}} \\
\cline{2-5}
  \bfseries  Approach & $T_D$ & $T_M$ & $D$  & $L$ \\
\cline{1-5}
Inter-copy C/R & - & ++ & ~+~ &  ~+~ \\
Pre-copy C/R & + & $\circ$ & $\circ$ & $\circ$ \\
Post-copy C/R & $\circ$ & ~+~ & $\circ$ & $\circ$ \\
Hybrid C/R & + & ~+~ & -& $\circ$ \\
Parallel Process Migration & ++ & $\circ$ & - & - \\
\cline{1-5}
\end{tabular}
\label{tab:comparison}
\end{center}
\end{table}

First, inter-copy \gls{c/r} is most simple and requires only a small time for the whole migration process, since the process is immediately stopped, migrated, and resumed. As a result of this, inter-copy \gls{c/r} is the most efficient method. However, the fact that the process downtime is equally high as the migration time, makes this approach not applicable for industrial control processes.
Since the pre-copy approach already transfers process states to the migration destination before the process is checkpointed, the process downtime is decreased. This leads to a longer overall migration time, a higher data volume that is transmitted, and a higher load on source and destination system.
In contrast, the post-copy method copies the process states in the end of the migration procedure. This approach is especially suitable if the migration time should be minimized, but inter-copy approach cannot be used due to the very high downtime. We believe that there are certain use cases, where this method is a good tradeoff, even if the process downtime is still too high for industrial control processes.
Hybrid \gls{c/r} combines both aforementioned approaches, but the advantages of this approach become only visible if the processes and corresponding sizes exceed certain limits. As there is a trend to use a lot of microservices instead of huge monolithic processes, this approach is not beneficial compared to either pre-copy or post-copy tactic if only small processes should be migrated. Moreover, it is not always trivial to determine which parts of the memory are urgently needed and which are not. 
As already mentioned, a reduced process downtime is the major goal of \gls{ppm} approach. Even if all other metrics are worse compared to all other approaches this is most suitable for the live migration of industrial control services. Thus, this approach serves as basis for several investigations \cite{govindaraj2018container,icit2021}. 

\section{Related Work}%
\label{sec:Related Work}
Live migration is already investigated for nearly two decades \cite{clark2005live}. The authors in \cite{torre2021benchmarking}, for example, prove that pre-copy tactic leads to a reduced process downtime compared to inter-copy \gls{c/r}. Moreover, the benefit is even high enough that the process downtime of migrating an \gls{vm} with pre-copy is lower than migrating a container with inter-copy, even though containers have much lower overhead compared to \glspl{vm}, being part of the research in \cite{7095802,indin2020,10.1145/2851613.2851737,7164727}.

Further, the authors in \cite{govindaraj2018container} and \cite{icit2021} focus on live migration of industrial control services. Even if both investigations use \gls{ppm} for live migration, \cite{govindaraj2018container} transfers the complete memory from the source to the target system, while \cite{icit2021} applies a more user-centric approach where only state-variables are transferred that were defined by the user beforehand. This makes the approach super fast, efficient, and platform independent, but requires knowledge of the control process. On the other hand, the approach of \cite{govindaraj2018container} does not require insights of the control process itself, but is a major achievement compared to inter-copy, pre-copy, post-copy, and hybrid \gls{c/r}. The only drawback, however, is that the process is still checkpointed. This means that it is intentionally stopped before the last memory is transmitted to the target system and resumed.

\section{Novel Concept}%
\label{sec:Novel Concept}
In order to further minimize the downtime a novel concept is proposed in this section.

\begin{figure}[tb]
\centering
  \includegraphics[width=\columnwidth]{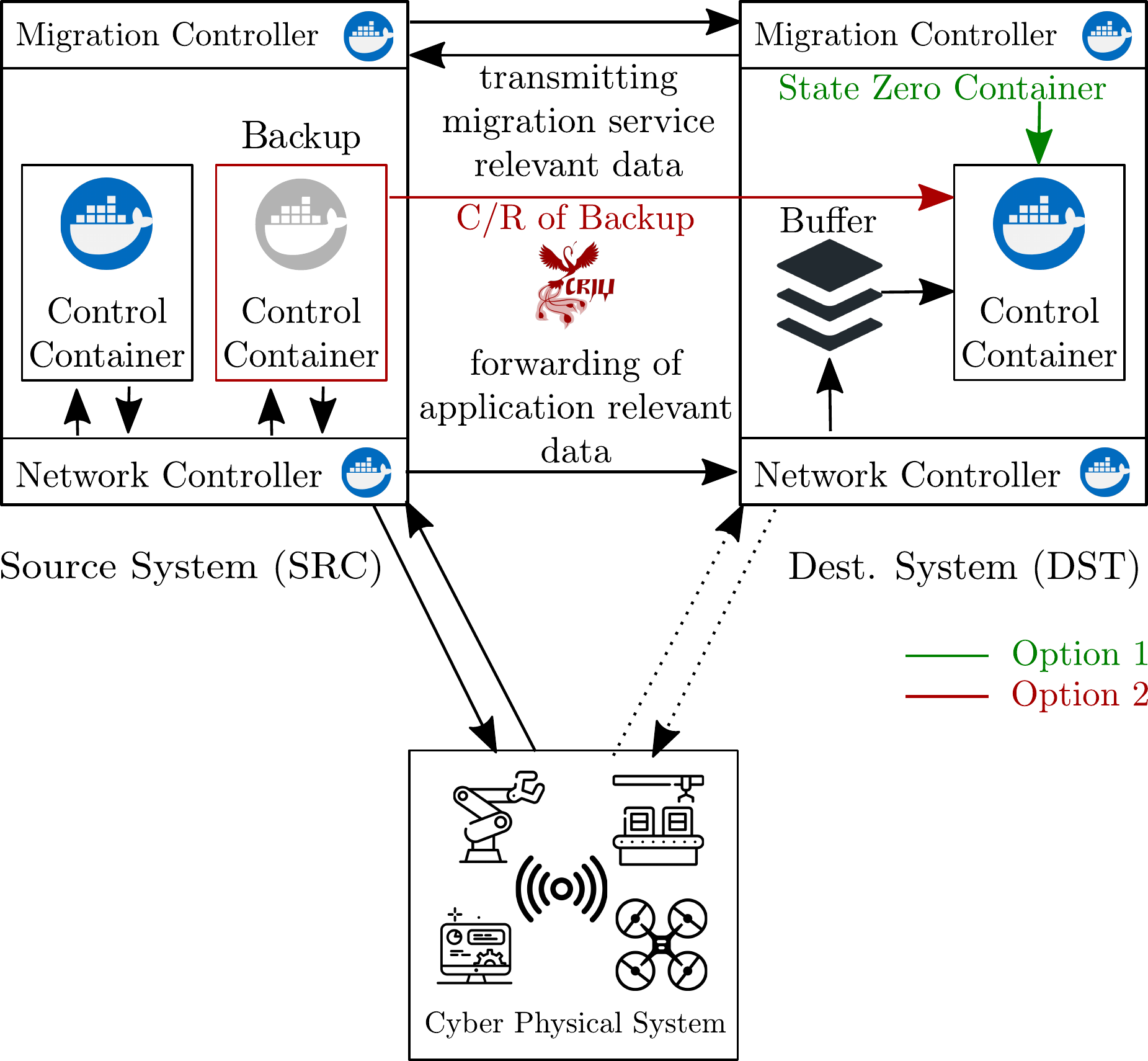} 
  \caption{Novel concept for live migration of industrial control processes with focus of minimized downtime.}
  \label{fig:concept}
\end{figure}

Here, two different hosts are assumed, where one of the hosts is the source ($SRC$) and the other the destination ($DST$) of the live migration, as shown in Fig. \ref{fig:concept}.
 Additionally, both hosts provide the compute resources $r_{SRC}$ and $r_{DST}$. Moreover, a migration controller, a network controller and a set of so-called control containers run on both of the devices. For the control containers, a resource consumption of $c_1,c_2,...,c_n$ is assumed.

Further, at least one \gls{cps} that can be any industrial device, sends data to a control container that is located on $SRC$. The control process itself can be described by a state vector $\underline{x}(t)$ with $n$ states, denoted by $x_1(t), ... , x_n(t)$ and an output vector  $\underline{y}(t)$ with $k$ outputs denoted by $y_1(t), ..., y_k(t)$, respectively.

The convergence criteria can be described by deviation $\underline{\delta}_y$ of the output vectors of $SRC$, denoted by $y_{a,1}, ...,y_{a,k}$ and $DST$, denoted by $y_{b,1}, ...,y_{b,k}$. Only if all $\delta_{y,i} = 0 $ or near sufficient $\delta_{y,i} \leq \eta_{y,i}$ a handover is initiated: 

\begin{equation} \label{eqn:output_dev}
\begin{split}
\underline{\delta}_y(t) =
\underline{y}_a(t) - \underline{y}_b(t)  =
\begin{pmatrix}
y_{a,1}(t) - y_{b,1}(t) \\
y_{a,2}(t) - y_{b,2}(t) \\
\vdots \\
y_{a,k}(t) - y_{b,k}(t)
\end{pmatrix} 
=
\begin{pmatrix}
\delta_{y,1}(t) \\
\delta_{y,2}(t) \\
\vdots \\
\delta_{y,k}(t)
\end{pmatrix}
\end{split}    
\end{equation}


The live migration process is described in Algorithms \ref{alg:migration1} \& \ref{alg:migration2}.

\begin{algorithm}
\caption{Migration process \textit{Option~1}}\label{alg:migration1}
\begin{algorithmic}[1]
\Require Migration request, Migration target $DST$
\Ensure $ \sum_{k=0}^{n}{c_k}+c_{n+1} < r_{DST}$
\State Connection setup b/w $SRC$ and $DST$
\State Start packet buffer on $DST$
\State Forward input data packets to $DST$
\State  Start state zero container on $DST$
\State Forward output data $\underline{y}$ to $DST$
\While{$\delta_y,i \nleq \eta_{y,i}, \forall i $} 
\State Send $\underline{x}$ to $DST$
    \While{packet buffer $>$ 0} 
    \State Read next packet
    \EndWhile  
\EndWhile  
\State  Initiate handover \Comment{Success}
\end{algorithmic}
\end{algorithm}

\begin{algorithm}
\caption{Migration process \textit{Option~2}}\label{alg:migration2}
\begin{algorithmic}[1]
\Require Migration request, Migration target  $DST$
\Ensure $ \sum_{k=0}^{n}{c_k}+c_{n+1} < r_{DST}$
\State Connection setup b/w $SRC$ and $DST$
\State Start packet buffer on $DST$
\State Forward input data packets to $DST$
\State Stop redundant control container on $SRC$
\State Checkpoint stopped control container on $SRC$
\State Transmit and restart control container on $DST$
\State Forward output data $\underline{y}$ to $DST$
\While{packet buffer $>$ 0} 
\State Read next packet
\EndWhile  
\If{$\delta_y,i\leq \eta_{y,i}, \forall i $}
\State  Initiate handover \Comment{Success}
\ElsIf{$\delta_y,i \nleq \eta_{y,i}, \forall i $}
\State Restart migration process \Comment{Error}
\EndIf  
\end{algorithmic}
\end{algorithm}

If a live migration from $SRC$ to $DST$  is triggered, it first has to be checked if an additional control container with the compute resources $c_{n+1}$ can be deployed on the target, or if the targets resources $r_{DST}$ are exceeded. If this is ensured, the connection between $SRC$ and $DST$ is established and a packet buffer is started on $DST$ by the migration controller. Further, all incoming packets that are addressed to the specific container on $SRC$ are copied, timestamped, and redirected to $DST$. These packets are cached in the packet buffer.

Next, a control container is started on $SRC$ by its migration controller. Dependant on the requirements of the maximum allowed migration time, two options are possible:
\begin{enumerate}
    \item Start of a state zero container and transfer of the state map $\underline{x}(t)$
    \item Start of a backup container at the beginning of the control process and \gls{c/r} to $DST$
\end{enumerate}

While \textit{Option~1} has a lower migration time, \textit{Option~2} can be applied non-invasive, which means that no process knowledge is required using this option. This is possible, since an identical container is started at the same starting point as the first container and receives all packets that the main control container receives from the beginning. Thus, any \gls{c/r} approach can be used to migrate this container from $SRC$ to $DST$. 
 
After the two control container instances are running on $SRC$ to $DST$ the output data packets that contain $\underline{y}$ of $SRC$ are forwarded to $DST$.

Then, for \textit{Option~1}, the state map $\underline{x}$ of $SRC$ is forwarded to $DST$. Subsequently both the state map and all data packets from the buffer with a timestamp equal to or greater than that of the state map are loaded and replayed in sequence. Afterwards it is checked if the pre-defined convergence criteria of Eq. \ref{eqn:output_dev} are fulfilled. 

On the other hand, for \textit{Option~2}, no state map $\underline{x}$ is required. Hence, all data packets from the buffer with a timestamp equal to or greater than that of the checkpoint are loaded and replayed in sequence and the convergence criteria are checked. If $\delta_y,i\leq \eta_{y,i}, \forall i $ the handover is initiated by the migration controller.




Even if the second option increases the migration time, it can be advantageous to have at least one backup container running in case of any error that can occur. In a realistic scenario, it makes also sense to deploy this backup container on another physical system than the main container. This makes the system more robust against hardware errors and faults in the communication link. If the backup has been proactively deployed to a potential target system for live migration and that system is selected as destination, the process itself does not need to be migrated.

\section{Testbed and Evaluation}%
\label{sec:Testbed and Evaluation}
In this section, our concept is evaluated. Therefore, a testbed consisting of real hardware  (Sec. \ref{subsec:testbed}) as well as an exemplary process (Sec. \ref{subsec: example process})  are proposed and the planned measurements are introduced (Sec. \ref{subsec:measurements}). Afterwards, the measurements are evaluated (Sec. \ref{subsec:evaluation}).

\subsection{Testbed}
\label{subsec:testbed}
In order to implement our concept in a realistic setup several hardware components are required, which are listed in Tab. \ref{tab:hardware}.

\begin{table}[tb]
\caption{Hardware configurations}
\begin{center}
\begin{tabular*}{\columnwidth}{|c|c|p{0.45\columnwidth}|}
\cline{1-3} 
\textbf{\textit{Equipment}} & \textbf{\textit{QTY}} & \textbf{\textit{Specification}}\\
\cline{1-3} 
Mini PC & 3 & Intel Core i7-8809G, 2x16 GB DDR4, Intel i210-AT \& i219-LM Gibgabit NICs, Debian 10.9.0 64 bit, \linebreak Linux 4.19.0-16-rt  \\
\cline{1-3} 
Network Switch & 1 & 8-Port Gigabit Ethernet Switch\\
\cline{1-3}
TSN Evaluation Kit & 1 & RAPID-TSNEK-V0001, IEEE~802.1AS-REV \\
\cline{1-3} 
\end{tabular*}
\label{tab:hardware}
\end{center}
\end{table}


First, two processing nodes are required that serve as source and target system for the migration process. Here, small sized mini PCs were used, because they could easily be integrated in existing industrial plants and serve as potential edge servers. Further, mini PCs are very often used for similar investigations. Thus a good comparability between the results is given. Additionally, both mini PCs are running on similar Linux OS and \gls{rt} Kernel. The latter is extremely important to achieve the required determinism of \gls{rt} applications. In order to perform the planned measurements of $T_M$ and $T_D$ for both options of our concept,  \textit{Docker} and  \textit{\gls{criu}} have been installed on the hosts. The latter is a software tool, widely used for inter-copy \gls{c/r} of both container and \glspl{vm}.

Moreover, both hosts have two  \glspl{nic}. This allows for the physical separation of \gls{rt} network that is required for the manufacturing process itself and a non \gls{rt} network for the integration in the factory network that provides management and orchestration interfaces and possibly also internet connection. In this work, the second network interface was used for the connection to a time master that synchronizes the clocks of both hosts, using \textit{linuxptp} open source software module. Here it was already shown that the synchronization error is $\ll$1µs \cite{icit2021}. Thus, this error can be neglected. 
Furthermore, both hosts are connected to each other and the \gls{cps} via a \gls{cots} network switch. A similar mini PC is used as \gls{cps}.

\subsection{Example Process}
\label{subsec: example process}
\begin{figure}[tb]
\centering
  \includegraphics[scale=1.2]{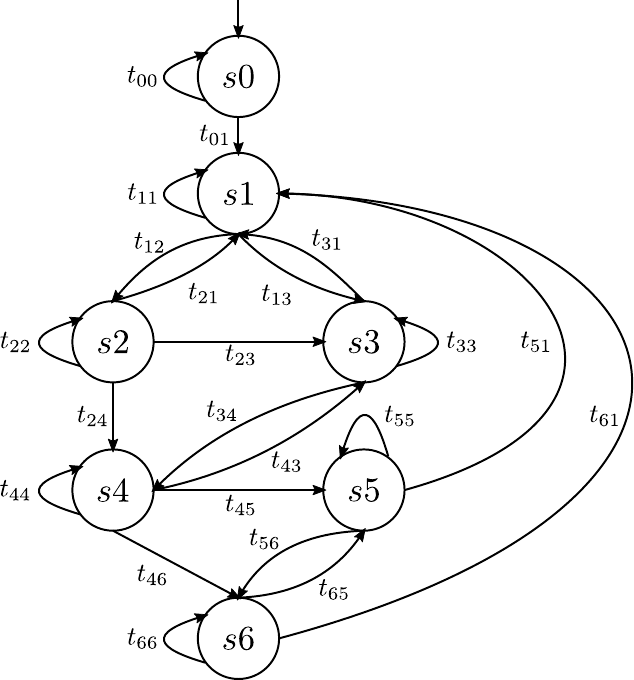} 
  \caption{Exemplary process including states and transitions.}
  \label{fig:example process}
\end{figure}

That the proposed concept can be tested properly, an exemplary process is specified (see Fig.~\ref{fig:example process}). Thus, a Markov chain is used that consists of the states $s_m$, with $m = 0, ..., 6$ and the transitions $t_{ij}$, where $i$ is the current state, and $j$ is the next state. Further, the state is updated once every 5 milliseconds and the probabilities for the transitions that are possible in the corresponding state are equally distributed. The transition probabilities are depicted in Eq.~\ref{eqn:output_dev}.

\begin{equation} \label{eqn:output_dev}
P =
\begin{pmatrix}
0.5 &  0.5   &  0  & 0   & 0 &  0 &  0 \\
0  & 0.\bar{3} &   0.\bar{3}  &  0.\bar{3}  & 0  & 0 &  0 \\
0 &  0.25  & 0.25  & 0.25  & 0.25 &  0 &  0 \\
0 &  0.\bar{3}  & 0  & 0.\bar{3}  & 0.\bar{3} &  0 &  0 \\
0 &  0  & 0  & 0.25 & 0.25 &  0.25 & 0.25 \\
0 &  0.\bar{3}  & 0  & 0  & 0 &  0.\bar{3} &  0.\bar{3} \\
0 &  0.\bar{3}  & 0  & 0  & 0 &  0.\bar{3} &  0.\bar{3} \\
\end{pmatrix}
\end{equation}

\subsection{Measurements}
\label{subsec:measurements}

In order to implement the novel concept in real industrial applications, the performance is decisive. Especially, if the process downtime $T_D$ is too high, a live migration under normal operation of the plant is not possible. To be clear, the process downtime has to be at least smaller as the update interval $T_U$, in which the \gls{cps} requires novel control information. Thus, the following condition has to hold:

\begin{equation}
    T_D \leq T_U
    \label{eq:1}
\end{equation}

Moreover, the duration of the whole migration process is relevant, since it specifies the dynamic of applications that require process migrations. Thus, if the migration time $T_M$ is very high, this could be a show stopper, even if Eq. \ref{eq:1} is fulfilled. Hence, Eq. \ref{eq:2} and Eq. \ref{eq:3} express the migration times for \textit{Option~1} and \textit{Option~2}, respectively:

\begin{equation}
    T_{M|Option~1} =  T_{image|prep} + T_{create} + T_{start} + T_D
    \label{eq:2}
\end{equation}

\begin{equation}
    T_{M|Option~2} =  T_{M|Option~1}  + T_{c/r}
    \label{eq:3}
\end{equation}

Looking into Eq. \ref{eq:2}, it can be seen that the migration time is the sum of image preparation  $T_{image|prep}$, creation and start of the container including the process to be migrated ($T_{create}$, $T_{start}$), as well as the process downtime ($T_D$). In order to describe the migration time for \textit{Option~2}, the time that is consumed by \gls{criu} for applying \gls{c/r} mechanism of the backup container ($T_{c/r}$) has to be added (see Eq. \ref{eq:3}). Furthermore, for the image preparation process there are two basic scenarios considered:
\begin{enumerate}
    \item The image is already located on $DST$. 
    \item The image has to be transferred to $DST$.
\end{enumerate}
If the image is already located on $DST$, it can be assumed that $T_{image|prep}=0$ since no data has to be transferred over the network. However, maintaining the images of all services on each possible node would result in a high space consumption on the disks of the nodes as well as to inconsistency if any change to an image is done. Thus, this scenario is only realistic, if the corresponding process was already located on this hardware node. This could be the case, if an \gls{agv}, such as a drone, is operating at the boundary of two factory floors, and the process needs to be live migrated at high frequency. On the other hand, in the second scenario, the image has to be transferred to $DST$. Here, it is assumed that the image is already located at least inside the factory network. This is realistic since the container including the process is already running on $SRC$ and the corresponding image had to be build in advance. Here, two more possibilities are included:
\begin{enumerate}
    \item The image is located at any data server in the central cloud of the factory. 
    \item The image is located at $SRC$.
\end{enumerate}

Since the provision of all images of the processes that are running on one system might not be efficient in terms of space consumption of the disk, it could be beneficial to build and maintain the image only on a centralized data server that is part of the factory cloud. Therefore, the time $T_{pull}$ for the pull of the image from the factory cloud to the $DST$ is assumed. 

If the image is available at $SRC$, it is beneficial to send the image directly to $DST$. Here, two widely used protocols for file transfer, namely \gls{scp} and \gls{rcp}, are integrated. These protocols offer simple security mechanisms by the requirement of user credentials and encryption. Since this results in an overhead that could significantly increase the migration time, a customized server-client communication for exchanging the image using \gls{tcp} was also realized.

\subsection{Evaluation}
\label{subsec:evaluation}

\begin{figure}[tb]
\resizebox{\columnwidth}{!}{%
%
%
\definecolor{mycolor1}{rgb}{0.00000,0.3,0.6}
\begin{tikzpicture}

\begin{axis}[%
width=3.5in,
height=0.5in,
scale only axis,
unbounded coords=jump,
xlabel style={font=\color{white!15!black}},
xlabel={$T_D$ [ms]},
ymin=0.7,
ymax=1.6,
ytick={1},
yticklabels={\empty},
ylabel style={font=\color{white!15!black}},
ylabel={~},
axis background/.style={fill=white},
xmajorgrids,
ymajorgrids,
]
\addplot [color=black, dashed, forget plot]
  table[row sep=crcr]{%
0.284   1\\
0.215    1\\
};
\addplot [color=black, dashed, forget plot]
  table[row sep=crcr]{%
0.15	1\\
0.049  1\\
};
\addplot [color=black, forget plot]
  table[row sep=crcr]{%
0.284 	0.925\\
0.284 	1.075\\
};
\addplot [color=black, forget plot]
  table[row sep=crcr]{%
0.049	0.925\\
0.049	1.075\\
};
\addplot [color=black, forget plot]
  table[row sep=crcr]{%
0.15	0.85\\
0.215	0.85\\
0.215	1.15\\
0.15	1.15\\
0.15	0.85\\
};
\addplot [color=red, forget plot]
  table[row sep=crcr]{%
0.184	0.85\\
0.184	1.15\\
};
\addplot [color=blue, only marks, mark=x, mark options={solid, draw=blue}, forget plot]
  table[row sep=crcr]{%
0.21	1\\
};
\addplot [color=black, only marks, mark=o, mark options={solid, draw=black}, forget plot]
  table[row sep=crcr]{%
0.313 1\\
0.328 1\\
0.346 1\\
0.361 1\\
0.37  1\\
0.444 1\\
0.557 1\\
0.683 1\\
0.868 1\\
  };
 
\node[color=blue] at (0.26,1.25) {\footnotesize{$0.21$~ms}};
\node[color=red] at (0.15,1.4) {\footnotesize{$0.18$~ms}};
\end{axis}
\end{tikzpicture}%
}
	\caption{Identified process downtime $T_D$ during a live migration process using our concept.}
\label{fig:downtime}
\end{figure}
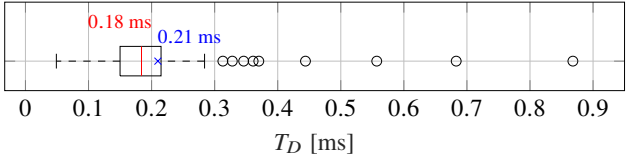

First, the results on the identified process downtime, which are depicted in Fig. \ref{fig:downtime}, are analysed.  
It can be seen that $T_D$ lies consequently under 1ms. Looking deeper into the data set, it can be concluded that both median and average value of $T_D$ lie in the area of 0.2ms. This means that also use cases can supported that have very stringent requirements on latency and determinism. To be clear, use cases that require update times of 0.2ms and more can be live migrated without any visible effects.

\begin{figure}[tb!]
\begin{tikzpicture}
    \node[anchor=south] at (0,0) { \includegraphics[width=0.7\columnwidth, trim = 80 89 75 80, clip]{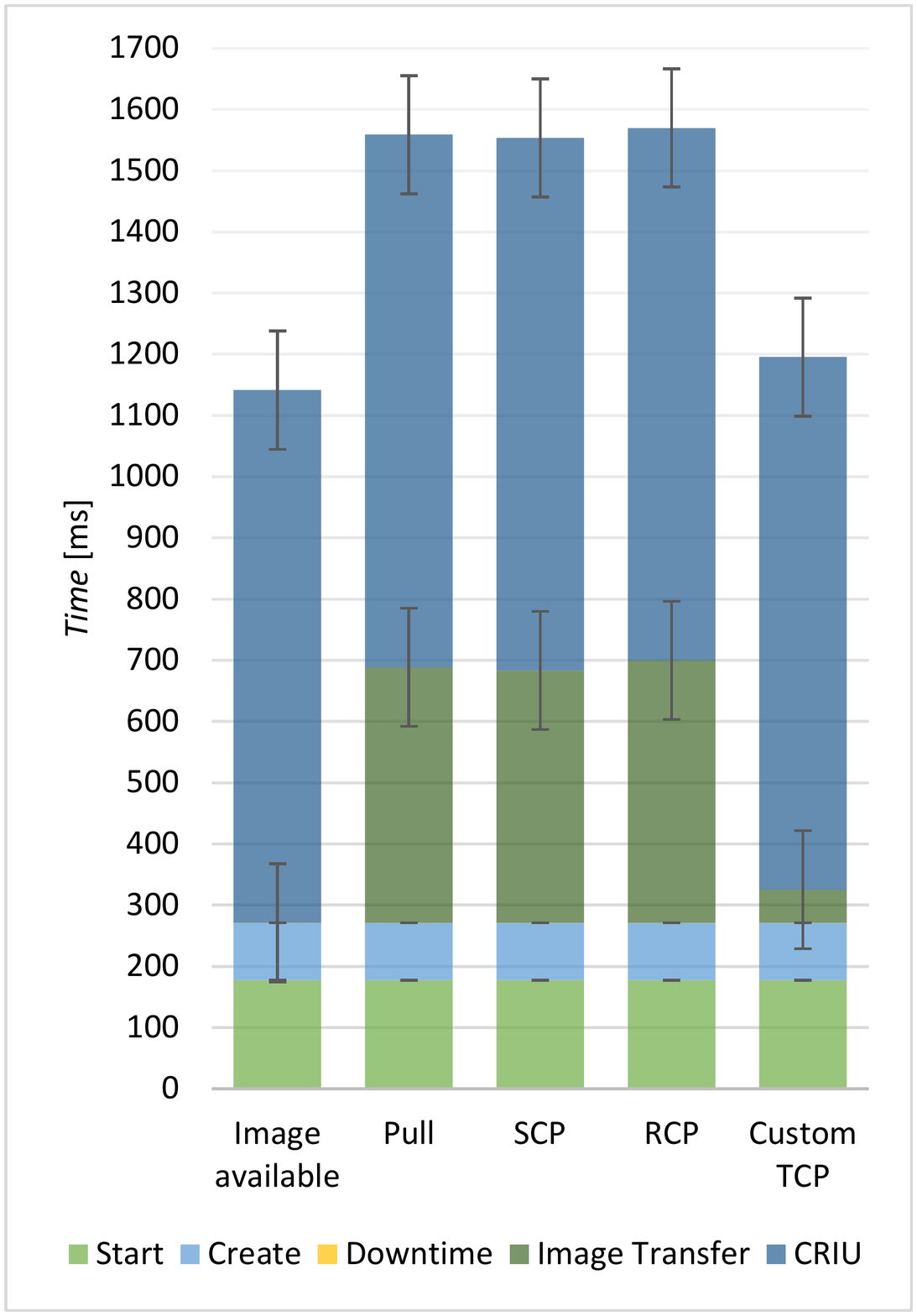}};
    \draw [color=blue](-.68in,1.085in) -- (1.25in, 1.085in);
    \draw [color=blue](.9in,1.185in) -- (1.35in, 1.185in);
    \draw [color=blue](.1in,1.84in) -- (1.45in, 1.84in);
    \draw [color=blue](-.3in,1.86in) -- (1.55in, 1.86in);
    \draw [color=blue](.5in,1.88in) -- (1.65in, 1.88in);
    
    \draw [](-.75in,.58in) -- (2.25in,.58in);
    
    \draw [latex-latex,color=blue](1.2in,.58in) -- (1.2in,1.085in);
    \draw [latex-latex,color=blue](1.3in,.58in) -- (1.3in,1.185in);
    \draw [latex-latex,color=blue](1.4in,.58in) -- (1.4in,1.84in);
    \draw [latex-latex,color=blue](1.5in,.58in) -- (1.5in,1.858in);
    \draw [latex-latex,color=blue](1.6in,.58in) -- (1.6in,1.88in);

    \node[color=blue] at (1.45in,2in) {\footnotesize{\textit{Option~1}}};
    
    \draw [latex-latex,color=red](1.8in,.58in) -- (1.8in,2.69in);
    \draw [latex-latex,color=red](1.9in,.58in) -- (1.9in,2.79in);
    \draw [latex-latex,color=red](2in,.58in) -- (2in,3.445in);
    \draw [latex-latex,color=red](2.1in,.58in) -- (2.1in,3.465in);
    \draw [latex-latex,color=red](2.2in,.58in) -- (2.2in,3.485in);
    
    \draw [color=red](-.68in,2.69in) -- (1.85in, 2.69in);
    \draw [color=red](.9in,2.79in) -- (1.95in, 2.79in);
    \draw [color=red](.1in,3.445in) -- (2.05in, 3.445in);
    \draw [color=red](-.3in,3.465in) -- (2.15in, 3.465in);
    \draw [color=red](.5in,3.485in) -- (2.25in, 3.485in);

    \node[color=red] at (1.7in,3.6in) {\footnotesize{\textit{Option~2}}};
\end{tikzpicture}
  \caption{Migration times $T_M$ for \textit{Option~1} and \textit{Option~2} based on different image transfer methods.}
  \label{fig:readings}
\end{figure}

Next, Fig.~\ref{fig:readings} displays the second metric $T_M$. 
The readings of $T_M$ indicate that it lies consequently below 1s for \textit{Option~1} and below 2s for \textit{Option~2}. This means that a live migration can be performed more than once per second for \textit{Option~1} and more than once every 2 seconds for \textit{Option~2}. Moreover, the biggest fraction for the migration time of \textit{Option~2} is the time consumed for applying \gls{c/r} method. This confirms our effort in developing a novel concept for live migration, since most existing approaches are still using this approach for live migration. Thus, $T_{CRIU}$ would be a fraction of the process downtime that is 1000 times higher compared to our approach. Even if $T_{CRIU}$ is only a fraction of $T_M$ in our approach, the migration time without using \textit{Option~2} is significantly lower. Thus, if possible, \textit{Option~1} should be applied, if a high migration dynamic is required. Further, the overhead of \gls{scp} and \gls{rcp} can be observed that is comparable to a pull of the image located at a central data server. Reasons for this overhead were already mentioned before. Hence, it can be concluded that a customized \gls{tcp} server-client communication is the preferred method. Using this method, a migration time of 325ms can be reached that is only about 15\% higher compared to the scenario, where the image is directly available on the target system. Here, a migration time of $\approx$270ms can be realized. Thus, live migrations of processes can be performed up to 3-4 times per second, using our optimized approach.     

\section{Conclusion}%
\label{sec:Conclusion}
In this paper, a novel concept for a downtime optimized live migration that is targeted for industrial \gls{rt} services was investigated. Therefore,the state-of-the-art was introduced and the different approaches based on specified metrics that are most relevant were analysed . Additionally, related work was proposed and existing challenges were discussed. In order to improve these challenges a novel concept was proposed and evaluated based on a testbed and exemplary process. The results clearly show that live migration using this concept only causes a process downtime in sub-milliseconds range. Hence, the proposed concept is able to fulfill mission-critical use cases that have very high demands on determinism and latency.

Furthermore, the migration time lies consequently below 1 second for the live migration of time invariant state-zero process controllers and below 2 seconds for using \gls{c/r} on a backup container. This means that a live migration of a mission-critical service can performed at least once per second, or every 2 seconds, respectively. Using a simplified transfer of the image, or is the image already available on the target node, the migration time can be further reduced below 300ms. Thus, a live migration can be performed multiple times per second. This allows for highly flexible control services that are prerequisite for realizing Industry 4.0 scenarios.



\printbibliography%

@InProceedings{indin2020,
  author       = {M. Gundall and D. Reti and H.~D. Schotten},
  booktitle    = {2020 IEEE 18th International Conference on Industrial Informatics (INDIN)},
  title        = {Application of Virtualization Technologies in Novel Industrial Automation: Catalyst or Show-Stopper?},
  year         = {2020},
  organization = {IEEE},
  pages        = {283--290},
  volume       = {1},
}

@Article{access2021,
  author  = {M. Gundall and M. {Strufe} and H. D. {Schotten} and P. {Rost} and others },
  journal = {IEEE Access},
  title   = {Introduction of a 5G-Enabled Architecture for the Realization of Industry 4.0 Use Cases},
  year    = {2021},
  pages   = {25508-25521},
  volume  = {9},
  doi     = {10.1109/ACCESS.2021.3057675},
}

@InProceedings{etfa2018,
  author    = {M. Gundall and J. Schneider and H. D. Schotten and M. Aleksy and others },
  booktitle = {2018 IEEE 23rd International Conference on Emerging Technologies and Factory Automation (ETFA)},
  title     = {5G as Enabler for Industrie 4.0 Use Cases: Challenges and Concepts},
  year      = {2018},
  month     = {Sep.},
  pages     = {1401-1408},
  volume    = {1},
  doi       = {10.1109/ETFA.2018.8502649},
  issn      = {1946-0759},
}

@InProceedings{icit2021,
  author    = {M. {Gundall} and C. {Glas} and H. D. {Schotten}},
  booktitle = {2021 22nd IEEE International Conference on Industrial Technology (ICIT)},
  title     = {Feasibility Study on Virtual Process Controllers as Basis for Future Industrial Automation Systems},
  year      = {2021},
}

@Article{lasi2014industry,
  author    = {H. Lasi and P. Fettke and H. Kemper and T. Feld and others },%M. Hoffmann}

@InProceedings{Mobilkom2021,
  author    = {M. Gundall and J. Stegmann and C. Huber and H. D. {Schotten}},
  booktitle = {Mobile Communication - Technologies and Applications; 25. ITG-Symposium},
  title     = {Towards Organic 6G Networks: Virtualization and Live Migration of Core Network Functions},
  year      = {2021},
  month     = {October},
  pages     = {1-5},
  issn      = {null},
}

@InProceedings{precopy1,
  author    = {Nie, Huqing and Li, Peng and Xu, He and Dong, Lu and others },%Song, Jinquan and Wang, Ruchuan}

@Article{performance1,
  author  = {Puliafito, Carlo and Vallati, Carlo and Mingozzi, Enzo and Merlino, Giovanni and others },% Longo, Francesco and Puliafito, Antonio}

@InProceedings{parallel1,
  author    = {Ma, Lele and Yi, Shanhe and Li, Qun},
  booktitle = {Proceedings of the Second ACM/IEEE Symposium on Edge Computing},
  title     = {Efficient Service Handoff across Edge Servers via Docker Container Migration},
  year      = {2017},
  address   = {New York, NY, USA},
  publisher = {Association for Computing Machinery},
  series    = {SEC '17},
  doi       = {10.1145/3132211.3134460},
  isbn      = {9781450350877},
}

@Article{parallel2,
  author  = {Wang, Shangguang and Xu, Jinliang and Zhang, Ning and Liu, Yujiong},
  journal = {IEEE Access},
  title   = {A Survey on Service Migration in Mobile Edge Computing},
  year    = {2018},
  pages   = {23511--23528},
  volume  = {6},
  doi     = {10.1109/ACCESS.2018.2828102},
}

@Book{reber1,
  author      = {Adrian Reber},
  title       = {Process Migration in a Parallel Environment},
  address     = {Stuttgart},
  date        = {2016},
  doi         = {10.18419/opus-8791,},
  institution = {Institut f{\"u}r H{\"o}chleistungsrechnen},
}

@Article{goldschmidt2018container,
  author    = {Goldschmidt, T. and Hauck-Stattelmann, S. and Malakuti, S. and Gr{\"u}ner, S.},
  journal   = {Journal of Systems Architecture},
  title     = {Container-based architecture for flexible industrial control applications},
  year      = {2018},
  pages     = {28--36},
  volume    = {84},
  publisher = {Elsevier},
}

@InProceedings{8502526,
  author    = {S. Grüner and S. Malakuti and J. Schmitt and T. Terzimehic and others },%M. Wenger and H. Elfaham}

@article{hozdic2015smart,
  title={Smart factory for industry 4.0: A review},
  author={Hozdi{\'c}, Elvis},
  journal={International Journal of Modern Manufacturing Technologies},
  volume={7},
  number={1},
  pages={28--35},
  year={2015}
}

@inproceedings{torre2021benchmarking,
  title={Benchmarking Live Migration Performance Under Stressed Conditions},
  author={Torre, Roberto and Schmoll, Robert-Steve and Kemser, Florian and Salah, Hani and others },% Tsokalo, Ievgenii and Fitzek, Frank HP}

@inproceedings{govindaraj2018container,
  title={Container live migration for latency critical industrial applications on edge computing},
  author={Govindaraj, Keerthana and Artemenko, Alexander},
  booktitle={2018 IEEE 23rd International Conference on Emerging Technologies and Factory Automation (ETFA)},
  volume={1},
  pages={83--90},
  year={2018},
  organization={IEEE}
}

@InProceedings{10.1007/978-3-642-36883-7_19,
author="Al-Shaer, Ehab
and Duan, Qi
and Jafarian, Jafar Haadi",
editor="Keromytis, Angelos D.
and Di Pietro, Roberto",
title="Random Host Mutation for Moving Target Defense",
booktitle="Security and Privacy in Communication Networks",
year="2013",
publisher="Springer Berlin Heidelberg",
address="Berlin, Heidelberg",
pages="310--327",
isbn="978-3-642-36883-7"
}

@article{hines2009post,
  title={Post-copy live migration of virtual machines},
  author={Hines, Michael R and Deshpande, Umesh and Gopalan, Kartik},
  journal={ACM SIGOPS operating systems review},
  volume={43},
  number={3},
  pages={14--26},
  year={2009},
  publisher={ACM New York, NY, USA}
}

@inproceedings{strunk2012costs,
  title={Costs of virtual machine live migration: A survey},
  author={Strunk, Anja},
  booktitle={2012 IEEE Eighth World Congress on Services},
  pages={323--329},
  year={2012},
  organization={IEEE}
}

@inproceedings{10.1145/2851613.2851737,
author = {Moga, Alexandru and Sivanthi, Thanikesavan and Franke, Carsten},
title = {OS-Level Virtualization for Industrial Automation Systems: Are We There Yet?},
year = {2016},
isbn = {9781450337397},
publisher = {Association for Computing Machinery},
address = {New York, NY, USA},
url = {https://doi.org/10.1145/2851613.2851737},
doi = {10.1145/2851613.2851737},
booktitle = {Proceedings of the 31st Annual ACM Symposium on Applied Computing},
pages = {1838–1843},
numpages = {6},
location = {Pisa, Italy},
series = {SAC ’16}
}

@InProceedings{7095802,
  author    = {W. Felter and A. Ferreira and R. Rajamony and J. Rubio},
  title     = {An updated performance comparison of virtual machines and Linux containers},
  booktitle = {2015 IEEE International Symposium on Performance Analysis of Systems and Software (ISPASS)},
  year      = {2015},
  pages     = {171-172},
  month     = {March},
  doi       = {10.1109/ISPASS.2015.7095802},
}

@inproceedings{clark2005live,
  title={Live migration of virtual machines},
  author={Clark, Christopher and Fraser, Keir and Hand, Steven and Hansen, Jacob Gorm and others },%Jul, Eric and Limpach, Christian and Pratt, Ian and Warfield, Andrew}

@INPROCEEDINGS{7164727, author={A. M. {Joy}}, booktitle={2015 International Conference on Advances in Computer Engineering and Applications}, title={Performance comparison between Linux containers and virtual machines}, year={2015}, volume={}, number={}, pages={342-346},}

@article{jiang2021road,
  title={The road towards 6G: A comprehensive survey},
  author={Jiang, W. and Han, B. and Habibi, Mohammad Asif and Schotten, Hans Dieter},
  journal={IEEE Open Journal of the Communications Society},
  volume={2},
  pages={334--366},
  year={2021},
  publisher={IEEE}
}
%
%
\end{document}